\documentclass[superscriptaddress]{revtex4}
\usepackage{graphicx}
\usepackage{titlesec} % title format

\titleformat{\section}[hang]{\bf\fontfamily{lmss}\large}{}{}{} % title format
\titleformat{\subsection}[runin]{\bf\fontfamily{lmss}}{}{}{} % title format

\begin{document}
	
	\title{Multilayer network analysis of nuclear reactions}
	
	\author{Liang Zhu}
	\affiliation{Shanghai Institute of Applied Physics, Chinese Academy of Sciences, Shanghai 201800, China}
	\affiliation{University of Chinese Academy of Sciences, Beijing 100049, China}
	
	\author{Yu-Gang Ma\footnote{Correspondence and requests for materials should be addressed to Y.G.M. (email: ygma@sinap.ac.cn).}}
	\affiliation{Shanghai Institute of Applied Physics, Chinese Academy of Sciences, Shanghai 201800, China}
	\affiliation{ShanghaiTech University, Shanghai 200031, China}
	
	\author{Qu Chen}
	\affiliation{School of Information Science and Technology, East China Normal University, Shanghai 200241, China}
	\affiliation{Shanghai Key Laboratory of Multidimensional Information Processing, East China Normal University, Shanghai 200241, China}
	
	\author{Ding-Ding Han}
	\affiliation{School of Information Science and Technology, East China Normal University, Shanghai 200241, China}
	\affiliation{Shanghai Key Laboratory of Multidimensional Information Processing, East China Normal University, Shanghai 200241, China}

\begin{abstract}
The nuclear reaction network is usually studied via precise calculation of differential equation sets, and much research interest has been focused on the characteristics of nuclides, such as half-life and size limit. In this paper, however, we adopt the methods from both multilayer and reaction networks, and obtain a distinctive view by mapping all the nuclear reactions in JINA REACLIB database into a directed network with 4 layers: neutron, proton, $^4$He and the remainder. The layer names correspond to reaction types decided by the currency particles consumed. This combined approach reveals that, in the remainder layer, the $\beta$-stability has high correlation with node degree difference and overlapping coefficient. Moreover, when reaction rates are considered as node strength, we find that, at lower temperatures, nuclide half-life scales reciprocally with its out-strength. The connection between physical properties and topological characteristics may help to explore the boundary of the nuclide chart.
\end{abstract}

\maketitle

\section*{Introduction}

The universe synthesizes lighter particles into heavier ones, and the mechanism which remains a heated area of exploration, includes element nucleosynthesis \cite{burbidge1957synthesis} and nuclear processes \cite{schatz1998rp-process,arnould2003p-process,arnould2007r-process,kappeler2011s-process,ji2016r-process,martin2015impact,hotokezaka2015short-lived}. Much attention has been drawn on topics like the limits of nuclear binding and the island of stability \cite{erler2012limits,oganessian2012nuclei,qian2014half-lives}. Those studies help us to understand nuclear reactions taking place both in the distant galaxies and in the reactors on the earth. Detailed studies of nuclide consumption or production through reactions are required with the support of massive nuclear data. The JINA REACLIB database, maintained by the Joint Institute for Nuclear Astrophysics, is a well-known source of thermonuclear reaction rates. It aims to compile a complete set of nuclear reaction rates that is continuously updated and regularly snapshotted \cite{cyburt2010jina}. The dataset contains various kinds of nuclear reactions, parameters for calculating reaction rates, $Q$ values, \textit{etc.}, providing rich information for nuclear-related research directions.

Besides acquiring exact nuclide abundance by solving sets of time-dependent differential equations in the database, it is challenging to treat the nuclear reaction system as a complex network to explore its statistical characteristics. The basic idea is introduced from graph theory, which considers the interacting units in a system as nodes and the relationship between two units as an edge, thus the system can be studied as a graph (network). The topic of complex networks has achieved significant advances since the `small world' \cite{watts1998collective} and `scale-free' \cite{barasi1999emergence} characteristics were found prevalent in many real world systems such as social connections, the Internet and distributed infrastructures \cite{newman2010networks}. The structure and dynamics of networks mapped from those systems turn out to be distinct from that of regular or random networks \cite{boccaletti2006complex,barrat2008dynamical,cohen2010complex}, and complex networks outperform them in modeling real world systems. For example, the `scale-free' structure of the Internet, which is hierarchical with many hubs, explains how easy it is for viruses to propagate. These findings help us to understand the systems at more profound levels and have benefited researches in various areas\cite{costa2011analyzing}, hopefully including nuclear reactions.

An important branch of the studies on complex networks covered metabolic networks or chemical reaction networks. The topological scaling properties were first analyzed of 43 organisms from different kinds of life in a directed network approach and similarity was found within and outside biological systems \cite{jeong2000large-scale}. It is believed that `scale-free' was crucial for the robustness. Moreover, the chemistry of planetary atmospheres and the interstellar medium were compared, and it was demonstrated that only the network of the Earth, coupling with its biosphere, was scale-free \cite{sole2004large-scale}. Efforts were made for obtaining more results from different systems like interstellar chemistry \cite{wagner2001small,jolley2010network-theoretical,jolley2012topological}, trying to explain the origins of complexity in large systems or to guide experiments \cite{hirdt2013data}. A majority of the studies map their reaction systems into networks using substrate-product method, which treats reactants and products as nodes and connects them with edges if they are in the same reaction equation. In reference \cite{holme2009model}, several mapping methods were evaluated. It was concluded that substrate-product and substance networks could best preserve modularity information. Although effective to a certain degree, as further discussion was made on precision of the methods used \cite{arita2004metabolic,montanez2010when}, the methods for mapping reactions into networks still need improvement to better fit a specific system in order to avoid inaccuracy. Worth mentioning, the bipartite graph is also an efficient way to characterize reactions \cite{guimera2007module}. It maps the substance and reaction into two kinds of nodes in one network, but the analysis of bipartite graph is complicated. We believe the best way is to extract the most information with a simpler method.

In the present work, we study a directed weighted nuclear reaction network based on substrate-product method, including all nuclides and reactions from the JINA REACLIB database. Additionally, we label each edge with 4 types: neutron, proton, $^4$He and the remainder, hence a 4-layered network. The multilayer network provides more precise description of a system, without which one might draw misinforming conclusions \cite{kurant2006layered}. To efficiently analyze the multilayer network, frameworks and models are studied \cite{dedomenico2013mathematical,sanchez-garcia2014dimensionality,nicosia2013growing,dedomenico2015structural,gao2012networks,kim2013coevolution}, and metrics for multilayer structure are proposed such as edge overlap \cite{bianconi2013statistical,battiston2014structural}, various centralities \cite{cozzo2015structure,dedomenico2015ranking} and correlations \cite{nicosia2016measuring,gemmetto2015multiplexity,menichetti2014correlations}, adding a new dimension to complex networks research. Detailed reviews on multilayer network are given in reference \cite{boccaletti2014structure,kivela2014multilayer,lee2015towards}. According to the dependence between nodes in different layers \cite{buldyrev2010catastrophic,parshani2010interdependent,shao2011cascade}, there are generally 3 types of multilayer network: one-to-one, partially dependent and one-to-many. We focus on the one-to-one dependent situation, and the multilayer substrate-product method falls into the multidimensional network category, which is originated from colored edge network. There are by far some attempts on biomedicine in reaction/metabolic systems using multilayer method. Here, the advantage of multilayer network analysis helps us to find some structural characteristics of nuclides relating to their physical property, which is a possible path to address the nuclide stability problem that nuclear physics and astrophysics pay close attention to.

\section*{Results}

\subsection*{Multilayer reaction network}

The dataset used in this paper is the latest version of REACLIBV2.0 containing 8048 nuclides treated as nodes and 82851 reactions mapped into directed edges (pairs from each reactant to each product). Over 3000 nuclides, either exist naturally or are synthesized artificially, whose half-life ($t^h$) data are complemented from NuDat (http://www.nndc.bnl.gov/nudat2/), while the rest are from theoretical calculation. Take $^{40}$Ca as an example, the recorded reactions involving this nuclide are as follows:
\begin{eqnarray*}
	^{40}\textrm{Ca} &\rightleftharpoons& \textrm{n} + ^{39}\textrm{Ca}\\
	^{40}\textrm{Ca} &\rightleftharpoons& \textrm{p} + ^{39}\textrm{K}\\
	^{40}\textrm{Ca} &\rightleftharpoons& ^4\textrm{He} + ^{36}\textrm{Ar}\\
	\textrm{n} + ^{40}\textrm{Ca} &\rightleftharpoons& \textrm{p} + ^{40}\textrm{K}\\
	\textrm{p} + ^{40}\textrm{Ca} &\rightleftharpoons& \textrm{n} + ^{40}\textrm{Sc}\\
	\textrm{p} + ^{40}\textrm{Ca} &\rightleftharpoons& ^4\textrm{He} + ^{37}\textrm{K}\\
	^4\textrm{He} + ^{40}\textrm{Ca} &\rightleftharpoons& \textrm{n} + ^{43}\textrm{Ti}\\
	^4\textrm{He} + ^{40}\textrm{Ca} &\rightleftharpoons& \textrm{p} + ^{43}\textrm{Sc}\\
	^{41}\textrm{Ca} &\rightleftharpoons& \textrm{n} + ^{40}\textrm{Ca}\\
	^{41}\textrm{Sc} &\rightleftharpoons& \textrm{p} + ^{40}\textrm{Ca}\\
	^{44}\textrm{Ti} &\rightleftharpoons& ^4\textrm{He} + ^{40}\textrm{Ca}\\
	^4\textrm{He} + ^{37}\textrm{Ar} &\rightleftharpoons& \textrm{n} + ^{40}\textrm{Ca}\\
	^{40}\textrm{K} &\rightarrow& ^{40}\textrm{Ca}\\
	^{40}\textrm{Sc} &\rightarrow& ^{40}\textrm{Ca}\\
	^{41}\textrm{Ti} &\rightarrow& \textrm{p} + ^{40}\textrm{Ca}\\
	\textrm{p} + ^{43}\textrm{Sc} &\rightarrow& ^4\textrm{He} + ^{40}\textrm{Ca}\\
	^{43}\textrm{Cr} &\rightarrow& \textrm{p} + \textrm{p} + \textrm{p} + ^{40}\textrm{Ca}
\end{eqnarray*}

\begin{figure}[ht]
	\centering
	\includegraphics[width=0.8\linewidth]{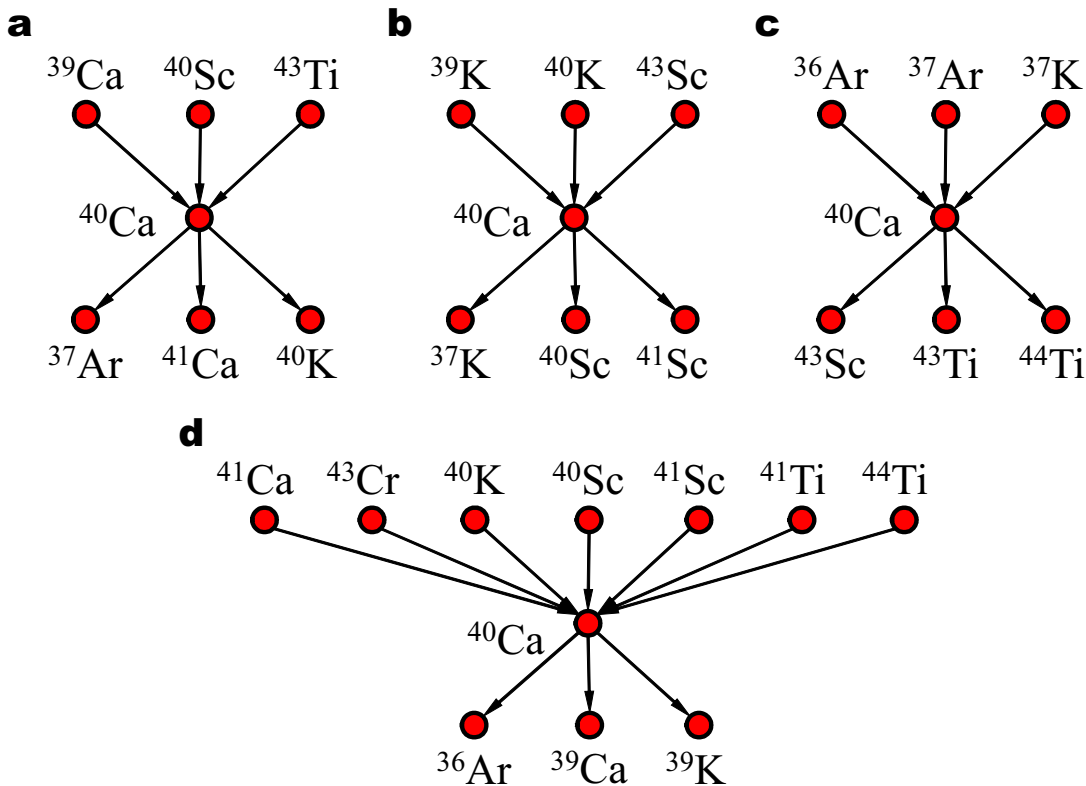}
	\caption{{\bf The topologies of $^{40}$Ca's 4 layers.} (a) n-layer, (b) p-layer, (c) h-layer, (d) r-layer. The network is constructed based on substrate-product method with edges labeled with currency particles in the reactants.}
	\label{fig:eg}
\end{figure}

\begin{figure}[ht]
	\centering
	\includegraphics[width=0.8\linewidth]{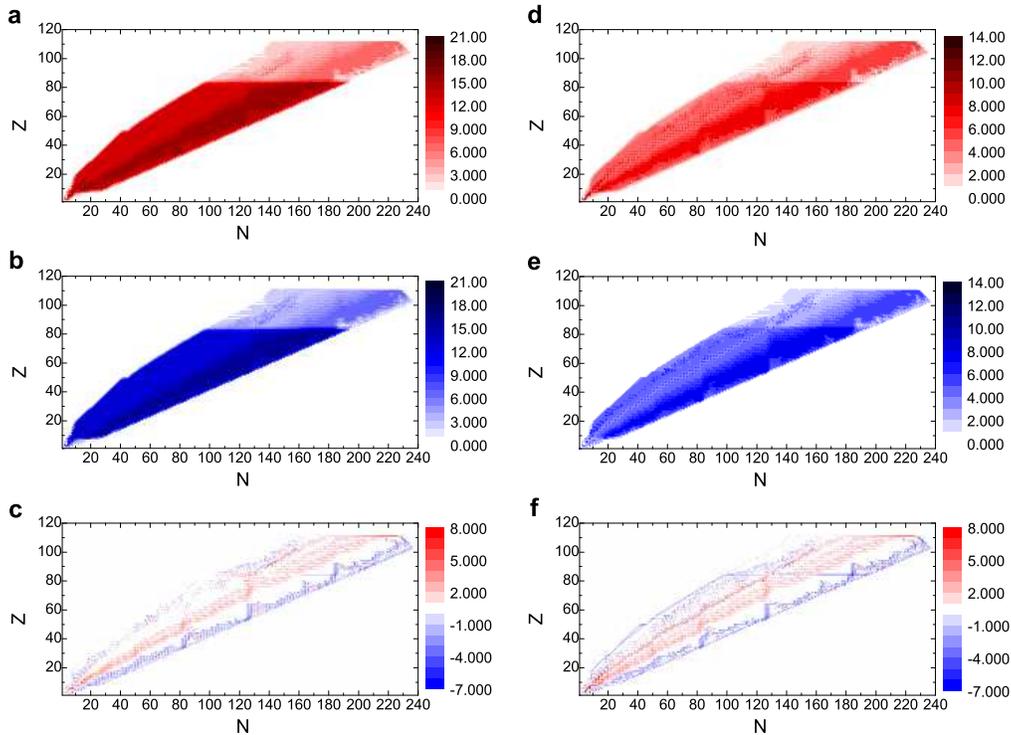}
	\caption{{\bf Degree distributions of aggregated network and r-layer on a $Z$-$N$ plane.} The in-degree and out-degree are calculated for each nuclide and the values are indicated by color, with x-axis being the number of neutrons ($N$) in that nuclide and y-axis being protons ($Z$). (a) In-degree, (b) Out-degree and (c) Degree difference are for the aggregated network. (d) In-degree, (e) Out-degree, (f) Degree difference are for the r-layer.}
	\label{fig:k2d}
\end{figure}

\begin{figure}[ht]
	\centering
	\includegraphics[width=0.8\linewidth]{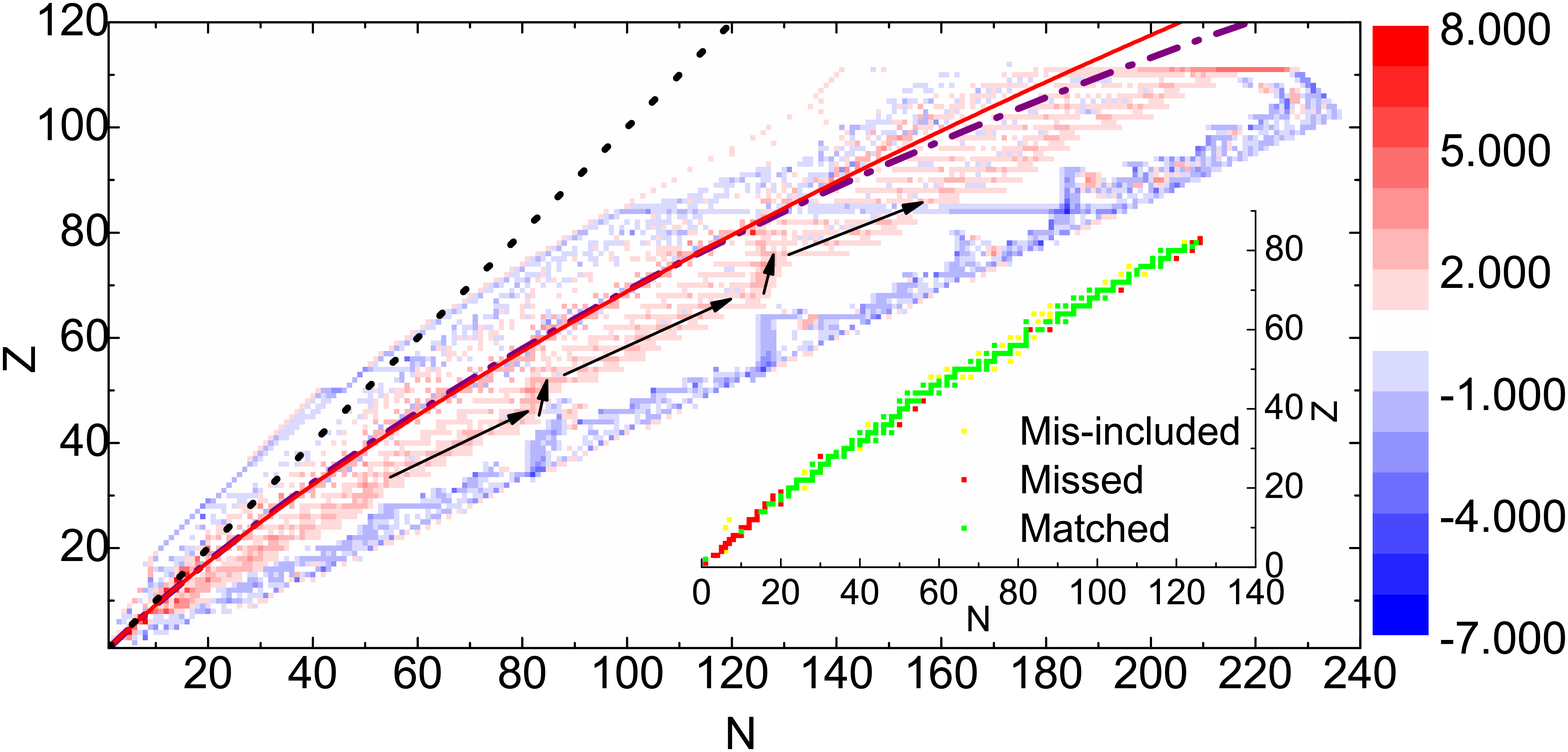}
	\caption{{\bf $\beta$-stability line fitting.} The valley of stability is fitted on the degree difference distribution of r-layer. The red solid line and the purple dash-dotted line are both empirical predictions of beta-stability lines where $Z=0.5A-0.3\times10^{-2}A^{5/3}$ and $Z=A(1.98+0.0155A^{2/3})^{-1}$.  The black solid lines with arrows could be able to outline the r-process. The black dotted line is auxiliary following $Z=N$. The inset demonstrates the fitting result of condition $(k_{i,\mathrm{in}}^{[r]}-k_{i,\mathrm{out}}^{[r]}=2)\bigcap(k_{i,\mathrm{in}}-k_{i,\mathrm{out}}=0)$. Green dots are 192 stable nuclides that match for the condition. Yellow dots are 31 unstable nuclides mis-included by the condition. Red dots are 40 stable nuclides missed by the condition. The ratio of correct prediction is relatively high.}
	\label{fig:betafit_r}
\end{figure}

According to the substrate-product method, for the first reaction of $^{40}$Ca, we have 4 edges: one from $^{40}$Ca to n and one from $^{40}$Ca to $^{39}Ca$, and vice versa because of the reverse reaction. However, as neutron, proton and $^4$He are the most needed particles in nuclear reactions, more than half of the reactions turn a nuclide into other nuclides by capturing or releasing one or more of these three particles, making their degrees $k$ (number of edges) close to $10^4$. If they are considered as normal nodes like other nuclides, whose degrees are smaller than $10^2$, their role of intermediary material will be lost in the topology and the degree distribution will be highly biased (see Supplementary Fig. S1 online). The nodes with extremely high degrees in reaction systems, often referred to as currency nodes, can make the network difficult to analyze, and they are usually deleted in reaction networks. Here, instead of completely deleting these currency nodes, we label an edge with one of the three particles (`n' for neutron, `p' for proton or `h' for $^4$He) when a nuclide needs this very particle to produce another nuclide. If none of the three particles are needed as reactant in a reaction, the edges from reactants to products will be labeled as `the remainder' or `r', which is not as simple as the situation for the other 3 layers and some information can be lost. Hence, the first reaction of $^{40}$Ca has only one r-edge from $^{40}$Ca to $^{39}$Ca and one n-edge from $^{39}$Ca to $^{40}$Ca. The corresponding topologies of $^{40}$Ca's 4 layers are shown in Figure \ref{fig:eg}.

Note that there are mainly 4 ways to map a reaction into a network: substrate-product network (connecting reactants to products), substrate-substrate network (reactants to reactants and products to products), substance network (the combination of the first and second methods), and reaction-reaction network (taking reactions as nodes and connecting 2 reactions if they share the same substance). Many reactions have multiple substrates while an edge in the network analysis usually connects only one pair of nodes, thus some topological structure could be lost when mapping a system to a network. However, in the case of nuclear reactions, the majority only have one or two reactants (see Supplementary Note online), so that the reactions are basically about one nuclide to another, making the information lost very limited. Also, if there are multiple products, some products can be decided following the conservation of mass (e.g. the 4$^{th}$ reaction of $^{40}$Ca, also written as $^{40}$Ca(n,p)$^{40}$K, the information about p is redundant). In this sense, the complete reaction information can be preserved. For other methods, the connection between reactants or products, as in substrate-substrate and substance network, could bring in misleading information, while reaction-reaction network loses too much. So the substrate-product network is our best choice. The topologies of the whole system can be found in Supplementary Fig. S2 online.

\subsection*{Topological characteristics and stable nuclides}

The network is usually studied via its adjacency matrix $A$. The matrix element $a_{ij}=1$ if there is an edge from node $i$ to $j$, otherwise $a_{ij}=0$. In a directed network, in-degree and out-degree are counted separately by summing up $A$ in different directions. As for multilayer network, we have $A^{[\alpha]}$ for layer $\alpha$, where, in our nuclear reaction network, $M=4$ and $\alpha\in\{n,p,h,r\}$. $A$ is denoted as the aggregated network of those 4 layers, $a_{ij}=1$ if $\sum\limits_{\alpha}{a_{ij}^{[\alpha]}}\ne0$. The degree distributions of aggregated network and r-layer on a $Z$-$N$ plane are shown in Figure \ref{fig:k2d}. The degrees of neutron-rich nuclides are mostly greater than those of proton-rich ones. While for n-, p- and h-layers, all the degree distributions have a peak at $k=3$, the networks are very homogeneous. We also notice that $k$ and $k_{nn}$ (averaging degrees of nearest neighbors) are highly correlated, which can be recognized as assortativity and has been found prevalent in many other networks. $k^{[r]}$ and $k_{nn}^{[r]}$ have a certain correlation, but there is little in the rest of 3 layers (see Supplementary Fig. S3-S6 and Supplementary Note online).

The abundant topological characteristics indicate there is rich information stored in the structure of the aggregated network and r-layer, so we explore them further by subtracting in-degree with out-degree as shown in the lower panels of Figure \ref{fig:k2d}. We can see that the degree difference of r-layer looks similar to the aggregated network, but there is a clear path-like positioned nodes whose in-degree is larger than out-degree (red dots), which are surrounded by scattered nodes that have smaller in-degree than out-degree (blue dots) along the path. On the other hand, the aggregated network does not possess this kind of pattern. As Figure \ref{fig:betafit_r} shows, which is an enlarged version of panel (f) in Figure \ref{fig:k2d}, this path fits the beta-stability line smoothly, and we would infer that this pattern describes most the positions of stable nuclides. Also, the remainder of the red part may be related to the r-process of nucleosynthesis.

\begin{figure}[ht]
	\centering
	\includegraphics[width=0.8\linewidth]{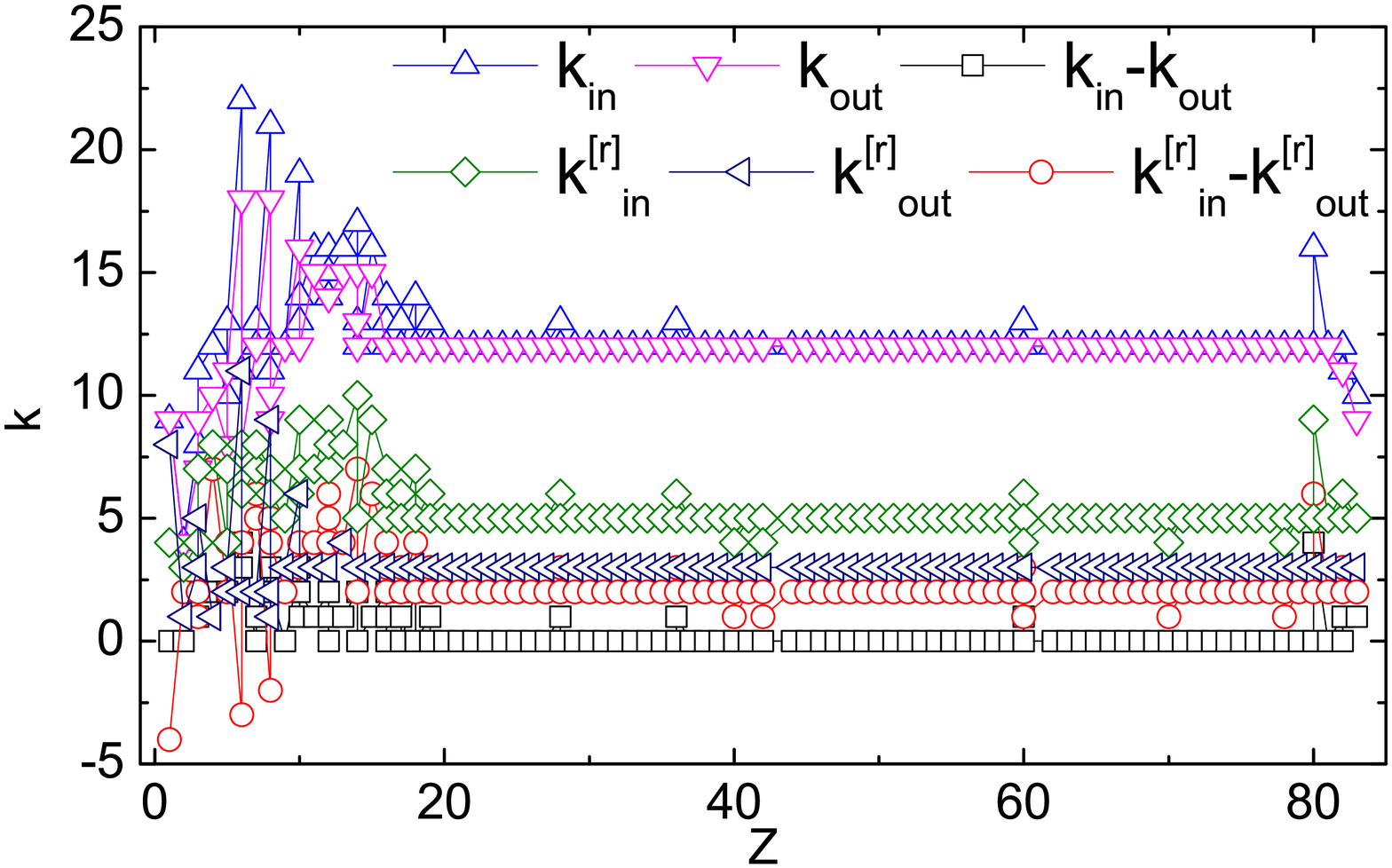}
	\caption{{\bf Degree difference over number of protons.} In the region of approximately $Z\ge15$, the in-degree and out-degree in the aggregated network are equal, while the difference is mostly 2 in the r-layer. For $Z<15$, the stable nuclides show topological complexity.}
	\label{fig:degreediff}
\end{figure}

\begin{figure}[ht]
	\centering
	\includegraphics[width=0.6\linewidth]{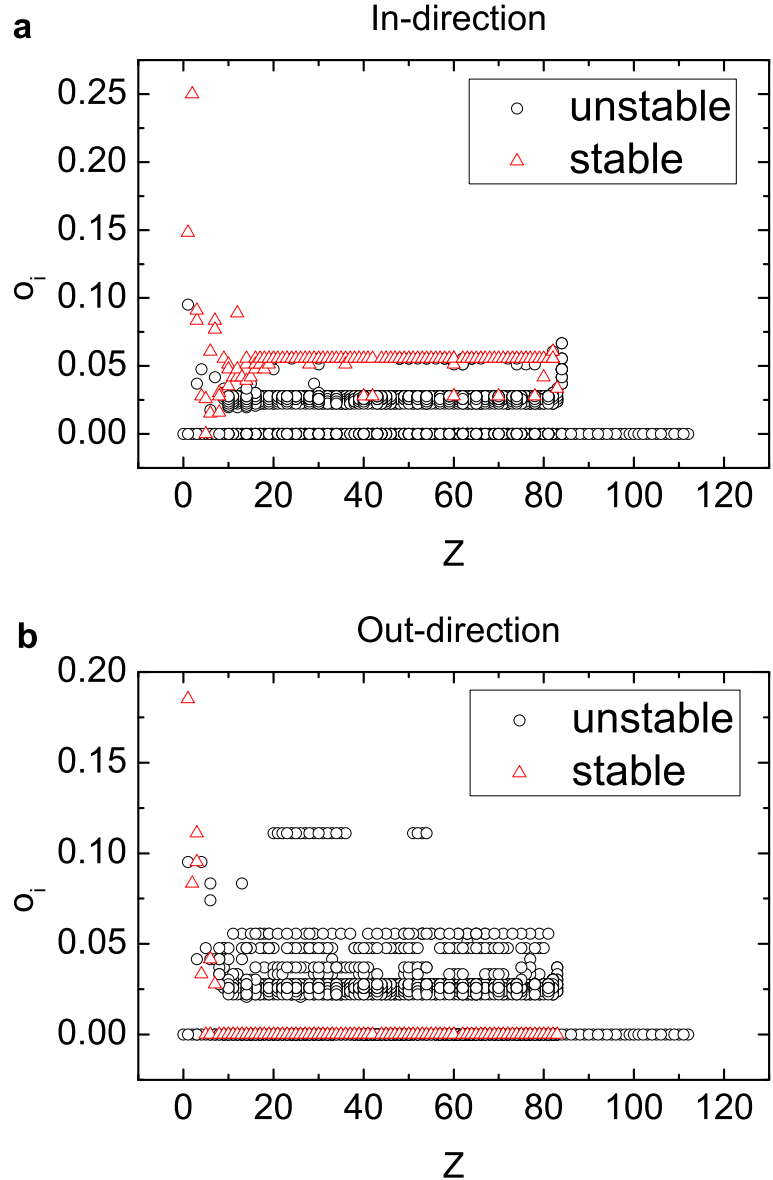}
	\caption{{\bf Node overlapping coefficient.} (a) $o_{i,\mathrm{in}}$ and (b) $o_{i,\mathrm{out}}$ of all nuclides are plotted over number of protons. The red triangles mark the stable nuclides which have high $o_{i,\mathrm{in}}$ and low $o_{i,\mathrm{out}}$.}
	\label{fig:nodeoverlap}
\end{figure}

To determine the characteristics of stable nuclides quantitatively, we compared the degree differences of the aggregated network and r-layer, and found that over 80\% of the stable nuclides follow the condition $(k_{i,\mathrm{in}}^{[r]}-k_{i,\mathrm{out}}^{[r]}=2)\bigcap(k_{i,\mathrm{in}}-k_{i,\mathrm{out}}=0)$ (see the inset of Figure \ref{fig:betafit_r}). As in-degree and out-degree correspond to the general ability of production and consumption of a nuclide, the aggregated network describes the total balance, so it is reasonable for $k_{i,\mathrm{in}}=k_{i,\mathrm{out}}$ to be one of the stability conditions. While in r-layer, where the n, p, $^4$He reactions are excluded, the condition of $k_{i,\mathrm{in}}^{[r]}>k_{i,\mathrm{out}}^{[r]}$ could result that the stable nuclides can be accumulated. As Figure \ref{fig:degreediff} shows, this condition does not apply to the stable nuclides with number of protons smaller than 15, which is probably because the area has more longer edges according to our measure. Thus we can conclude that for the $Z\geq15$ part, the topology is more regular, while the rest part has more complexity. Furthermore, the stable condition indicates that the degree differences in other 3 layers are not the same for in and out situations, so that the difference in r-layer can be compensated, eliminating the difference in the aggregated network. We checked this inference via node overlapping coefficient as depicted in Figure \ref{fig:nodeoverlap}. The stable nuclides do have much higher edge overlap in the in-direction than out-direction. In fact, $o_{i\in\{\mathrm{stable}\},\mathrm{in}}$ are mostly the highest among all the nuclides, and $o_{i\in\{\mathrm{stable}\},\mathrm{out}}$ are the lowest down to zero.

\begin{figure}[ht]
	\centering
	\includegraphics[width=0.6\linewidth]{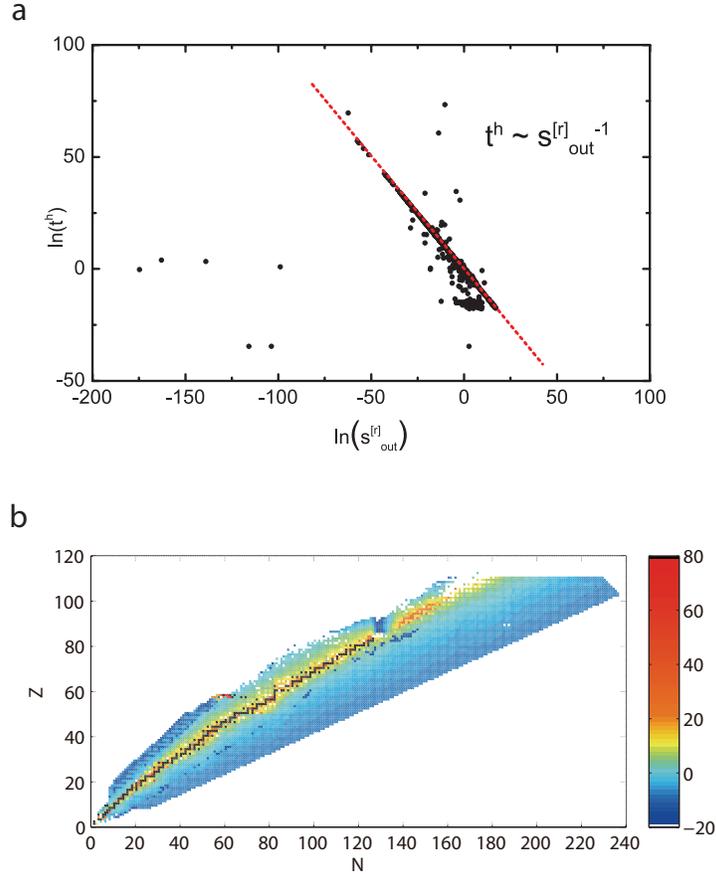}%eps}
	\vspace{-1.2cm}
	\caption{{\bf The relationship between half-life and out-strength of r-layer and a prediction.} (a) At $T_9=0.01$, the half-life of approximately 2500 out of 3000 nuclides scales reciprocally with node out-strength of r-layer. (b) A prediction of half-life based on Eq.~\ref{eq:halflife}. $\ln(t^h)$ of a nuclide is indicated by color, with x-axis being the number of neutrons ($N$) in that nuclide and y-axis being protons ($Z$).}
	\label{fig:halflife}
\end{figure}

\begin{figure}
	\centering
	\includegraphics[width=0.6\linewidth]{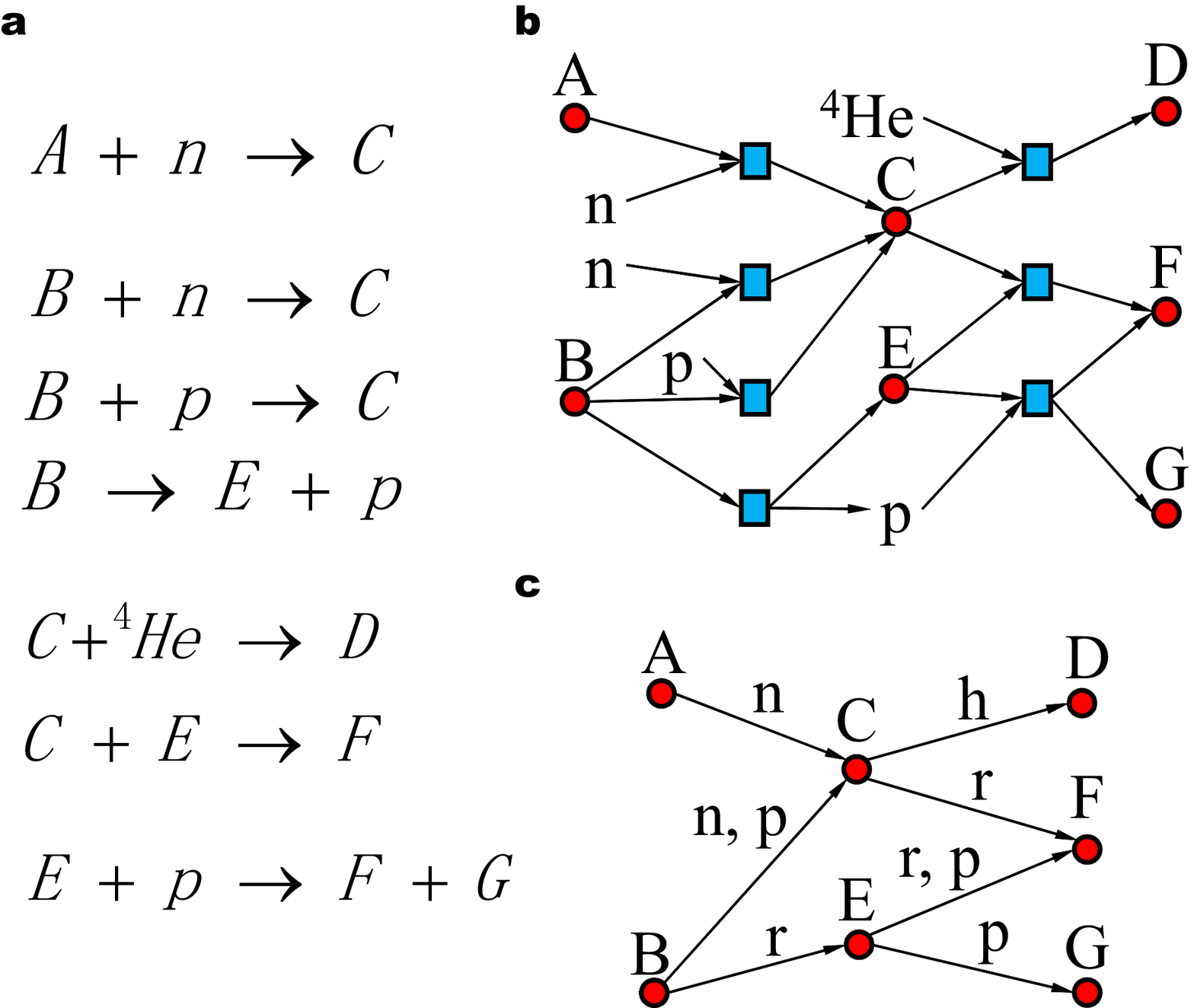}
	\caption{{\bf Mapping reactions into a network.} (a) pseudo equation sets including 4 kinds of reactions. (b) Mapping all the reactants and products into a network, similar to a bipartite graph, where a blue square stands for a reaction and a red circle for a nuclide. (c) The aggregated network. Edges are labeled according to the currency particles in the reactants.}\label{fig:mapping}
\end{figure}

\subsection*{Weighted analysis and nuclide half-life}

Networks without weights are usually called binary, which, in our case, only indicates the reactions that have the capability to take place. While the weights on the edges make the structure even more complicated and draw a clearer picture of the system. For each reaction, there are 7 parameters ($a_{i}$) for calculating reaction rates $\lambda$ using Equation \ref{eq:rate}, where $T_9$ is the temperature in Gigakelvin \cite{cyburt2010jina}. A larger $\lambda$ indicates higher probability for a reaction to occur, which can be mapped to the notion of edge weight in network theory, describing how easy it is to travel between nodes or how important the edge is.
\begin{equation}
\label{eq:rate}
\lambda =\exp [(a_0 +\sum\limits_{i=1}^5 {a_i T_9^{\frac{2i-5}{3}} } +a_6 \ln T_9 )]
\end{equation}

Although $k_{\mathrm{in}}$ and $k_{\mathrm{out}}$ are highly correlated in binary network (upper and middle panels of Figure \ref{fig:k2d}), $s_{\mathrm{in}}$ and $s_{\mathrm{out}}$ have various correlation under different temperature values as the reaction rate is a function of $T_9$ (see Supplementary Fig. S7 online). This could be the reason why a 4-layered network is able to lay foundation for all complex systems at atomic level with 3 layers being rather regular. The complexity comes from the diversity of edge weights. Of the temperature range we explored, it is observed that for $T_9\leq3$, a reciprocal relationship
\begin{equation}
\label{eq:halflife}
t_{i}^h\approx \frac{1}{e^{0.4}\cdot s_{i,out}^{[r]}}
\end{equation}
\noindent exists between $t^h$ and $s_{out}^{[r]}$ as in the upper panel of Figure \ref{fig:halflife}. Approximately 2500 out of 3000 nuclides with half-life values fit for Equation \ref{eq:halflife} if the temperature is low, where $t^h$ is in second. The proportion of fitted nuclides drops as $T_9$ increases. Those unfitted points could be caused by insufficient data to form the network. This result can help to predict the half-lives of unknown nuclides when their reaction rates with neighbors are calculated, as shown in the lower panel of Figure \ref{fig:halflife}. A comparison with half-life distribution plotted using Nudat data can be found in Supplementary Fig. S8 online. Although correlated to $s_{out}^{[r]}$ to some extent, $s_{in}^{[r]}$ does not have such relationship with $t^h$, nor strengths of other layers. The reason for this phenomenon can be explained that half-life is determined by how easy it is for a nuclide to convert into others, and r-layer contains mainly reactions that are classified as one nuclide to others without any currency particles. Also, it indicates that the life-time of a nuclide does not depend on its input channel by which a nuclide is formed, i.e. the life-time of a certain nucleus has no memory for the formation history of itself.

\section*{Discussion}

The whole set of nuclear reactions is one of the most fundamental systems in the world. The network constructed is quite different from previously studied ones. The average shortest distance is a bit longer than the network diameter, which indicates that it is not such a small world. The degree distribution is narrow with peaks, far from the power-law often observed. Three layers of the 4-layered network are quite regular. The nodes are spatially constrained, lacking enough long range edges to make the communication or propagation more efficient, yet they can generate much more complicated items like chemical molecules and biological tissues. The temperature-dependent strength may take the credit for the complexity. Despite the special structure, some interesting node characteristics are observed from the complex network point of view.

The multilayer network construction method is effective, even some of the products are deleted, we can still extract important information from the network. As stability is about the balance of in and out corresponding to production and consumption in reactions, the degree difference and node overlapping coefficient are able to identify the majority of stable nuclides, and the relationship between half-life and out-strength in r-layer are discovered. The results are obtained with the help of multidimensional method, linking nuclide characteristics with topological structures.

When we focus on a complicated reaction process (e.g. r-process is about continually capturing neutrons to produce heavier nuclides), it usually occurs only in a certain layer, even if the network structure is simple. Therefore, as we study nucleosynthesis and nuclear processes, based on the primitive results of nuclear reaction network analysis, our future work will focus on a better designed mapping method, the dynamics on the network and more detailed analysis of each layer.

\section*{Methods}

\subsection*{Mapping reactions into a network}

Figure \ref{fig:mapping} is a demonstration of our mapping method. Note that the number of edge types between two nodes can range from $0$ to the number of layers $M$.

\subsection*{Node degree and overlapping coefficient}

The in-degree and out-degree for node $i$ in layer $\alpha$ are defined as in Equation \ref{eq:k}. $\sum\limits_{\alpha}{a_{ij}^{[\alpha]}}$ is the number of different types of edges between node $i$ and $j$, called edge overlap, which is proposed in \cite{bianconi2013statistical}. An average edge overlap is defined in reference \cite{dedomenico2015structural}. To evaluate the overall overlap of a node in different layers, we define a normalized node overlapping coefficient $o_i$, as in Equation \ref{eq:overlap}, based on the edge overlap, where $k_i$ is the degree of node $i$ in the aggregated network.
\begin{equation}
\label{eq:k}
\begin{array}{l}
k_{i,in}^{[\alpha]} = \sum\limits_{j}{a_{ji}^{[\alpha]}} \\
k_{i,out}^{[\alpha]} = \sum\limits_{j}{a_{ij}^{[\alpha]}} \\
\end{array}
\end{equation}
\begin{equation}
\label{eq:overlap}
\begin{array}{l}
o_{i,in} = \frac{1}{k_{i,in}}\sum\limits_{j}{\frac{\sum\limits_{\alpha}{a_{ji}^{[\alpha]}}-1}{M-1}} \\
o_{i,out} = \frac{1}{k_{i,out}}\sum\limits_{j}{\frac{\sum\limits_{\alpha}{a_{ij}^{[\alpha]}}-1}{M-1}} \\
\end{array}
\end{equation}

\subsection*{Node strength}

We calculate the rates of each reaction and assign the values to the edges respectively as edge weight. The rate value ranges from approximately $10^{-300}$ to $10^{50}$ at $T_9=0.1$, and varies as $T_9$ grows from less than $0.01$ up to $10$. The weights of the two edges with opposite directions between a pair of nodes are usually not the same as the rates of a reaction and its reverse reaction often differ. For $T_9\ge0.01$, the calculation of reverse rates should be corrected by including partition functions, as they are calculated via detailed balance without partition functions in the dataset. To understand the characteristics of a node, we derive node strength $s_i$ by summing up edge weights from or to all the neighboring nodes as shown in Equation \ref{eq:s}.
\begin{equation}
\label{eq:s}
\begin{array}{l}
s_{i,in}^{[\alpha]} = \sum\limits_{j}{\lambda_{ji}\cdot a_{ji}^{[\alpha]}} \\
s_{i,out}^{[\alpha]} = \sum\limits_{j}{\lambda_{ij}\cdot a_{ij}^{[\alpha]}} \\
\end{array}
\end{equation}

\section*{Acknowledgements}

This work is supported in part by the Major State Basic Research Development Program in China under Contract No. 2014CB845400, the National Natural Science Foundation of China under contract Nos. 11421505 and 11520101004.

\section*{Author contributions statement}

Y.G.M presented the idea of this work. L.Z. and Y.G.M. designed the study and performed the analyses. L.Z. prepared the figures and drafted the manuscript. Q.C. and D.D.H. participated in the network analysis of the results. All authors approved the final version of the manuscript.

\section*{Additional information}
\textbf{Accession codes} (where applicable);\\
\textbf{Competing financial interests} The authors declare no competing financial interests.

\end{document}